\documentclass[conference]{IEEEtran}

\ifCLASSINFOpdf
\else
\fi
\usepackage{balance} 
\usepackage{graphics}
\usepackage{times}   
\usepackage{url}

\usepackage{graphicx}
\usepackage{caption}
\usepackage{subcaption}
\usepackage[cmex10]{amsmath}
\usepackage[normalem]{ulem}
\usepackage{cite}
\makeatletter
\def\url@leostyle{
  \@ifundefined{selectfont}{\def\UrlFont{\sf}}{\def\UrlFont{\small\bf\ttfamily}}}
\makeatother
\def\pprw{8.5in}
\def\pprh{11in}

\usepackage[pdftex]{hyperref}
\hypersetup{
pdftitle={Text Entry Method Affects Password Security},
pdfauthor={Yulong Yang, Janne Lindqvist, Antti Oulasvirta},
pdfkeywords={usable security},
bookmarksnumbered,
pdfstartview={FitH},
colorlinks,
citecolor=black,
filecolor=black,
linkcolor=black,
urlcolor=black,
breaklinks=true,
}

\usepackage{verbatim}

\usepackage{color}
\usepackage{multirow}
\newcommand{\textred}[1]{\textcolor{red}{#1}}
\ifx\noeditingmarks\undefined
   \newcommand{\pgwrapper}[2]{\textred{#1 #2}}
\else
   \newcommand{\pgwrapper}[2]{}
\fi

\IEEEoverridecommandlockouts

\begin{document}
\title{Text Entry Method Affects Password Security}
\author{\IEEEauthorblockN{Yulong Yang}
\IEEEauthorblockA{
Rutgers University\\
Email: yulong.yang@rutgers.edu}
\and
\IEEEauthorblockN{Janne Lindqvist}
\IEEEauthorblockA{
Rutgers University\\
Email: janne@winlab.rutgers.edu}
\and
\IEEEauthorblockN{Antti Oulasvirta}
\IEEEauthorblockA{MPI for Informatics and Saarland University\\
Email: oantti@mpi-inf.mpg.de}
}

\maketitle

\begin{abstract}
Text-based passwords continue to be the prime form of authentication to computer systems. Today, they are increasingly created and used with mobile text entry methods, such as touchscreens and mobile keyboards, in addition to traditional physical keyboards.
This raises a foundational question for usable security: whether text entry methods affect password generation and password security. 
This paper presents results from a between-group study with 63 participants, in which each group generated passwords for multiple virtual accounts using a different text entry method. Participants were also asked to recall their passwords afterwards. 
We applied analysis of structures and probabilities, with standard and recent security metrics and also performed cracking attacks on the collected data. 
The results show a significant effect of text entry methods on passwords. 
In particular, one of the experimental groups created passwords with significantly more lowercase letters per password than the control group ($t(60) = 2.99, p = 0.004$). The choices for character types in each group were also significantly different ($p=0.048, FET$).
Our cracking attacks consequently expose significantly different resistance across groups ($p=0.031, FET$) and text entry method vulnerabilities. 
Our findings contribute to the understanding of password security in the context of usable interfaces.
\end{abstract}

\section{Introduction}

Text-based passwords remain as the most prevalent method of authentication \cite{Herley:2012:RAA:2360743.2360824}. 
In addition to traditional computers such desktops and laptops, people increasingly generate and use passwords with a wide variety of mobile terminals, such as tablets and smartphones. These mobile terminals have very different text entry methods compared to traditional physical keyboards available for desktops and laptops. Mobile terminals are also replacing traditional computers in daily tasks. For example, Pew Research estimates that 21\% of all US adult cell phone owners use primarily their phone to access the web~\cite{cellinternet2013}.

In the present study, we examined whether the design of text entry methods affect the security of generated passwords. A \emph{text entry method} \cite{MacKenzie:2007:TES:1296062} consists of those physical (e.g.~form factor, display, sensor, etc.) and software (e.g.~virtual keyboard layout) aspects of an input device that are relevant when entering text.
The design of a text entry method determines how quickly and effortlessly a given character can be typed. 
Even small changes in how characters are displayed and organized can affect both typing performance~\cite{zhai2002performance} and visual search~\cite{norman1982alphabetic}. 
Differences in selection time should be pronounced for text entry methods with vastly different form factors, such as a smartphone using a virtual keyboard versus a laptop with a physical keyboard. Further, typing performance among entry methods can differ remarkably: experienced typists on physical keyboards reach more than 60 words per minute (wpm)~\cite{gentner1983acquisition}, whereas tablets and smartphones are in the range of 20 to 30 wpm~\cite{goel2012walktype, oulasvirta2013improving}. 
We hypothesized that differences in the design of text entry method might affect password generation.
Such an effect would depend on whether users are mindful of such design when they mentally generate password candidates. For example, for numeric 4-digit automatic teller machine PIN codes, it has been found that PINs are created to produce visual patterns on the keygrid \cite{bonneauPINs}.

In particular, we hypothesized that, depending on password generation strategy, entry methods can affect password security in a few ways: users may generate passwords by using the characters on the display as generation cues. More precisely, the difficulty to reach a character from the present view should affect the probability of its inclusion in a password. Consider for example the fact that common laptop keyboard has 47 characters and common touchscreen qwerty keyboard has only 26 characters on the first screen. Not only do these entry methods differ in the entry rate of single keys~\cite{azenkot2012touch, salthouse1986perceptual}, they also most likely differ in the visual overhead of locating keys. 
A common laptop keyboard shows each letter, number or special character on the physical key. The user needs just to press
e.g.~shift to access special characters.
In contrast, common virtual keyboards require users to first switch layouts and then visually
search in each layout to locate the desired key.
These differences could have prominence in the generated password. One should also see corresponding differences in the distribution of characters. For instance, numbers are not directly reachable without changing the screen in the common touchscreen qwerty keyboard on smartphones -- does this affect the generated passwords? 

To better understand if an effect of text entry method design occurs, we conducted a randomized controlled between-group experimental study with 63 participants. The study allowed us to compare passwords generated with different text entry methods. We assigned participants into three groups, each of which was provided a different text entry method we chose purposefully. Participants were first asked to generate three passwords for three clearly distinct virtual accounts (bank, email, online magazine). They were then immediately tested for their memory, and asked to do that again after at least 10 days. We chose three text entry methods that were different from each other in several ways but still resemble the main present-day platforms for text entry. Specifically, we chose the text entry methods so that the difficulty of reaching different keys increased in experimental groups compared to the control group.

We found that the basic structures and distributions of passwords were significantly different across groups. In particular, as the difficulty of reaching keys increased, participants chose characters from a smaller subset of characters to construct passwords. As a result, the experimental group with the most difficult method had much more passwords containing only lowercase letters compared to other groups. However, to our surprise, passwords across groups seemed to have similar scores when applied with standard password strength metrics (entropy~\cite{Shannon:2001:MTC:584091.584093, 394764}, NIST entropy~\cite{Burr:2011:SEA:2206278}) and a recently introduced Markov-model-based metric (adaptive password-strength meter~\cite{Castelluccia:2012:NDSS:markov}). However,  we found that passwords across groups showed statistically significant different resistance against password cracking attacks. This is important for password security. In particular, when a vulnerability exists in the password population for a text entry method, attackers can use method-specific attacks. This could be possible for example against sites that are known to be accessed mostly by mobile terminals. Therefore, we conclude that text entry methods affect password security. We believe our results further motivate studies of password security in the context of usable interfaces.

This paper makes following contributions:

1. Our results show that text entry methods affect the security of user-generated passwords. Specifically, passwords created with different methods show significant difference in resistance against password cracking attacks.

2. Our results show that the text entry methods affect how people compose their passwords. For example, participants in experimental group two focused a small set of characters compared to the experimental group one and the control group.

3. We report a comprehensive analysis of user-generated passwords in laboratory settings. In particular, we confirm previous research suggesting that entropy-based metrics should not be used as indicators of password strength. The present work also calls more attention to a recently proposed Markov-model-based password strength meter~\cite{Castelluccia:2012:NDSS:markov}, and in particular its applicability to small-scale password dataset analysis.

\section{Related Work}

In this section, we focus on what is known about how people generate text-based passwords. For an overview of research
for the past twelve years on studies  on graphical passwords, an alternative method of authentication,
we recommend a survey by Biddle et al. \cite{Biddle:2012:GPL:2333112.2333114}.

Passwords have been used in computer systems since 1960s  \cite{Wilkes:1975:TSC:540274},
and have been studied and criticized since at least 1970s \cite{Saltzer:1974:PCI:361011.361067}. A case
history of password security dates back in 1979  \cite{Morris:1979:PSC:359168.359172}.
Jakobsson et al.~\cite{benefitspasswords} have argued that we should understand passwords better and
block weak passwords.

Florencio et al.~\cite{Florencio:2007:LSW:1242572.1242661} studied people's web password habits, and found that people's passwords were generally of poor quality, they are re-used and forgotten frequently. Grawemeyer \cite{Grawemeyer:2011:UMM:1994007.1994160} conducted a diary study to understand how people use passwords, they found how people had different strategies for example for different accounts. Recently, Bonneau et al.~\cite{bonneauPINs} studied how people choose 4-digit PINs for banking cards. Common strategies included birth dates and trivial patterns based on the key displays. The reported presence of visual strategies in the 4-digit case supports our hypothesis that password generated with text input methods, too, may differ.

Researchers have studied how password creation policies affect password security. Yan et al.~\cite{Yan:2004:PMS:1024867.1025014} were among the first to study empirically how different password policies affect security and memorability of the passwords. They found that mnemonic phrases as passwords were as easy to remember as naive passwords. Later Kuo et al.~\cite{Kuo:2006:HSM:1143120.1143129} studied human-selection of mnemonic phrases and found that their 400,000-entry dictionary cracked 4\% of mnemonic passwords; in comparison, a standard dictionary with 1.2 million entries cracked 11\% of control passwords.

Popular websites (e.g.~Google, Twitter) have adopted password meters as real-time measures of password strength for users. Ur et al.~\cite{Ur:2012:YPM:2362793.2362798} found that stringently rated password meters led users to make significantly longer passwords that included more digits, symbols, and uppercase letters, and the passwords were also more resistant cracking algorithms. However, Egelman et al.~\cite{Egelman:2013:MPG:2470654.2481329} showed that password meters do not help much if people consider the accounts unimportant. Another large-scale study on the effect of password policy indicate that, given stronger password requirements to meet, although users were annoyed, they also believe that they are now more secure; in addition, use of dictionary words and names are still the most common strategies to create passwords \cite{Shay:2010:ESP:1837110.1837113}. However, Weir et al.~\cite{Weir:2010:TMP:1866307.1866327} claim that, passwords created under common password requirements, such as minimum length and different character set requirements, are still vulnerable to cracking attacks, due to the fact that people tend to pick the easiest passwords that meet the requirement. Mazurek et al.~\cite{Mazurek:2013:MPG:2508859.2516726} found when measuring password guessability for an entire university that students affiliated with the computer science school created 1.8 times stronger passwords than those who were affiliated with the business school.

Recently, Houshmand et al.~\cite{Houshmand:2012:BBP:2420950.2420966} implemented AMP, which is a probabilistic cracking system for suggesting better passwords.  Inglensant et al.~\cite{Inglesant:2010:TCU:1753326.1753384} studied the costs for organizations from unusable password policies. Brostoff et al.~\cite{BroSas2003} predicted that requests for password reminders could be reduced by up to 44\% by increasing the number of how many times users can try their passwords before accounts get locked from three to ten. Zviran et al.~\cite{Zviran:1999:PSE:1189462.1189470} looked at how password policies influence people's tendencies to write passwords down, finding that e.g.~commonly used passwords are not written down. 

Chiasson et al.~\cite{Chiasson:2009:MPI:1653662.1653722} conducted laboratory studies on how people recall multiple text-based passwords compared to multiple click-based graphical passwords (PassPoints \cite{Wiedenbeck:2005:PDL:1090412.1090418}). They found that the recall rates after two weeks were not statistically significant from each other. Fahl et al.~\cite{Fahl:2013:EVP:2501604.2501617} compared real passwords to those generated in an experiment, finding that about 30\% of subjects do not behave as they do in real life. However, the authors concluded that laboratory studies generally create useful data. 

Finally, few researchers have specifically look at helping people to create passwords on non-traditional text entry methods. Schaub et al.~\cite{Schaub:2012:PEU:2406367.2406384} studied the usability and shoulder surfing susceptibility of different smartphone platforms. Haque et al.~\cite{Haque:2013:PIT:2516760.2516767} have studied how to create better passwords on mobile devices. Jakobsson et al.~\cite{rethinkingpasswords} proposed fastwords, which relies on standard error-correcting features for users to create passphrases. Mannan et al.~\cite{mohammadboth} implemented ObPwd, a tool for managing passwords on mobiles. The users choose an object (e.g. picture), and ObPwd creates a corresponding password that will be used on website automatically by the ObPwd plugin.

\section{Method}

Our study randomly assigned participants into one control group and two experiment groups. They were provided with a different text entry method per group. Then they were asked to created passwords for three different virtual accounts. Participants were told that they should generate as good passwords as they normally would and their ability to recall them would be tested later. They were asked to recall created passwords immediately after a short distraction task, and to recall it again after a time duration at least 10 days. To collect more data, we used a mixed method approach: after producing passwords, all participants filled a questionnaire about subjective workload (NASA-TLX \cite{hart:tlx}) and answered interview questions about the tasks they performed.

We conducted the study in a laboratory in order to control for confounding factors. A controlled laboratory experiment allows for choosing the main factors to be considered, in our case the
text entry method. Recently, a study by Fahl et al.\cite{Fahl:2013:EVP:2501604.2501617} indicates that lab studies provide for significantly more helpful and realistic data compared to online studies.

Next, we describe our volunteer participants, our experiment design, apparatus, measurement and data analyze methods.

\subsection{Participants}
We recruited participants through fliers, mailing lists, and in person at cafeterias. Participants were required to be 18 years old or over and familiar with touchscreen devices. 
We recruited 63 participants in total, between the ages of 18 to 65 ($M = 27.2, SD = 9.9$). Among all participants, 24 were male and 39 were female. The educational background of the participants during the study were: 22 had a high school diploma, 23 had a Bachelor's degree, 16 had a graduate degree and two participants had other degrees. Participants had an average of 3.44 years of experience with touchscreen devices, and on average spent 3.70 hours a day using a touchscreen device and 7.41 hours online per day. 

All 63 participants completed session one of our study, and 57 of them returned and participated in session two as well.
As compensation, participants received \$30 for completing the whole study. They also participated in a raffle of three \$75 gift cards.

We recruited our participants in two batches, 33 in May and 30 in June and July 2013.
Between the beginning of the recruiting batches there was a time gap of 70 days. Also, the gap between the two sessions of the study varied. The mean time gap for the first volunteers was 14.53 (SD = 5.81) days and 29.52 (SD = 7.57) days for the second.

Our study was approved by the Institutional Review Board of our institutions.

\subsection{Experiment Design}

The experiment followed a 3 x 3 mixed design: text entry method type (3 levels) was controlled as a between-subject variable and account type (3 levels) as a within-subjects variable.

The first variable was the main variable we were interested: the different types of text entry method. We divided our participants into three groups based on this variable: control group, experiment groups one and two. The comparison of these three groups is the main focus in our analysis. The participants were randomly assigned into one of these three groups. The participants were unaware of the assignments or that other groups existed. The detailed explanation for the differences between the group setup is given in the next subsection.

The main reason we designed this variable to be between-group was to isolate its effect from any other undesired effects such as any possible confounding factors that would correlate with both the variable and the result. In contrast, a within-subject design would not be appropriate. With a within-subject design, we would ask each participant perform the same creation task with all methods. This could introduce potential random effects within each subject with respect to our main variable, leading to inaccurate data.

We also included the type of virtual accounts as a within-subject variable, to be nested within our between-group variable. With this nested structure, we accomplished two things. First, we increased the sample size, which ensured the power of our experiment design. Second, this enabled us to study any potential interaction effect of our main variable given the context of different types of accounts.

We note that a similar design has been used before by Chiasson et al.~\cite{Chiasson:2009:MPI:1653662.1653722} to compare how generating multiple passwords affect the ability to recall text-based and multiple click-based graphical passwords (PassPoints \cite{Wiedenbeck:2005:PDL:1090412.1090418}). Our work differs from theirs in that they focused on analysis of recall sessions, while we focused on password creation process and password security. In particular, our recall session was designed to motivate participants to create realistic passwords, thus increasing ecological validity; our main focus was on the passwords created in session one. In contrast, their work focused on the recall rate collected from recall session. Moreover, we analyzed our result with also the account type variable and its interaction between our main variable when applicable. We also used TLX forms to estimate workload.

\subsection{Apparatus}

Our main experiment setup was defined by the apparatus (text entry methods) each group used. 

\textit{Control group} 

We provided a common laptop (Macbook Pro 2012 with a 13'' display) as the device used in the control group.  We chose this because the physical keyboard on a common laptop is likely to be still the most common text entry method for password creation. This laptop had a common physical qwerty keyboard in English language.

\textit{Experiment group one} 

We provided a large tablet (Nexus 10 tablet with Android 4.2.2, 10.1'' touchscreen) as the device used in experiment group one. The keyboard on the tablet had the same layout and usability as a common qwerty touchscreen keyboard. Given that the tablet can be held in the hands in two ways, we asked the participants to keep it in the ``landscape'' mode when using it.

\textit{Experiment group two} 

We provided a regular-size smartphone (Galaxy Nexus smartphone with Android 4.2.2, 4.5'' touchscreen) as the device used in the experiment group two. The keyboard layout on the smartphone was chosen from several available designs for smartphone platforms. We had four layouts: lowercase letters (first layout), digits and some special symbols (second layout), uppercase letters (third layout) and more special symbols (fourth layout), as the same as a common qwerty touchscreen keyboard. If we define the reachability of each layout by the number of key presses one needs to reach that layout from the first layout, in a common qwerty touchscreen keyboard, the reachability of the four layouts are 0 (pop up immediately when activated), 1 (in most cases denoted as ``123"), 1 (``Shift" key) and 2 (one more press after reaching second layout), respectively. However, we built our keyboard so that the our layouts were ordered hierarchically and the reachability of them became 0, 1, 2 and 3, respectively. In such way, one has to reach the previous layout before reaching the next one. Compared with the common design, our uppercase letters and some special symbols are harder to reach.

The primary reason we chose our text entry methods as described was to differentiate each group in their difficulty of reaching keys during text entry. Method in group one was more difficult to reach digits and symbols compared with that of control group in that it did not include them in the first layout; method in group two was more difficult than that of group one was because (1). smaller screen to interact with, (2). more switches between layouts for different types of keys. Note all three apparatus still provided good usability.

The keyboard layout design of both group one and two are demonstrated in Figure~\ref{fig:customized_kbd_layout}. 

\begin{figure*}
\includegraphics[width=\textwidth]{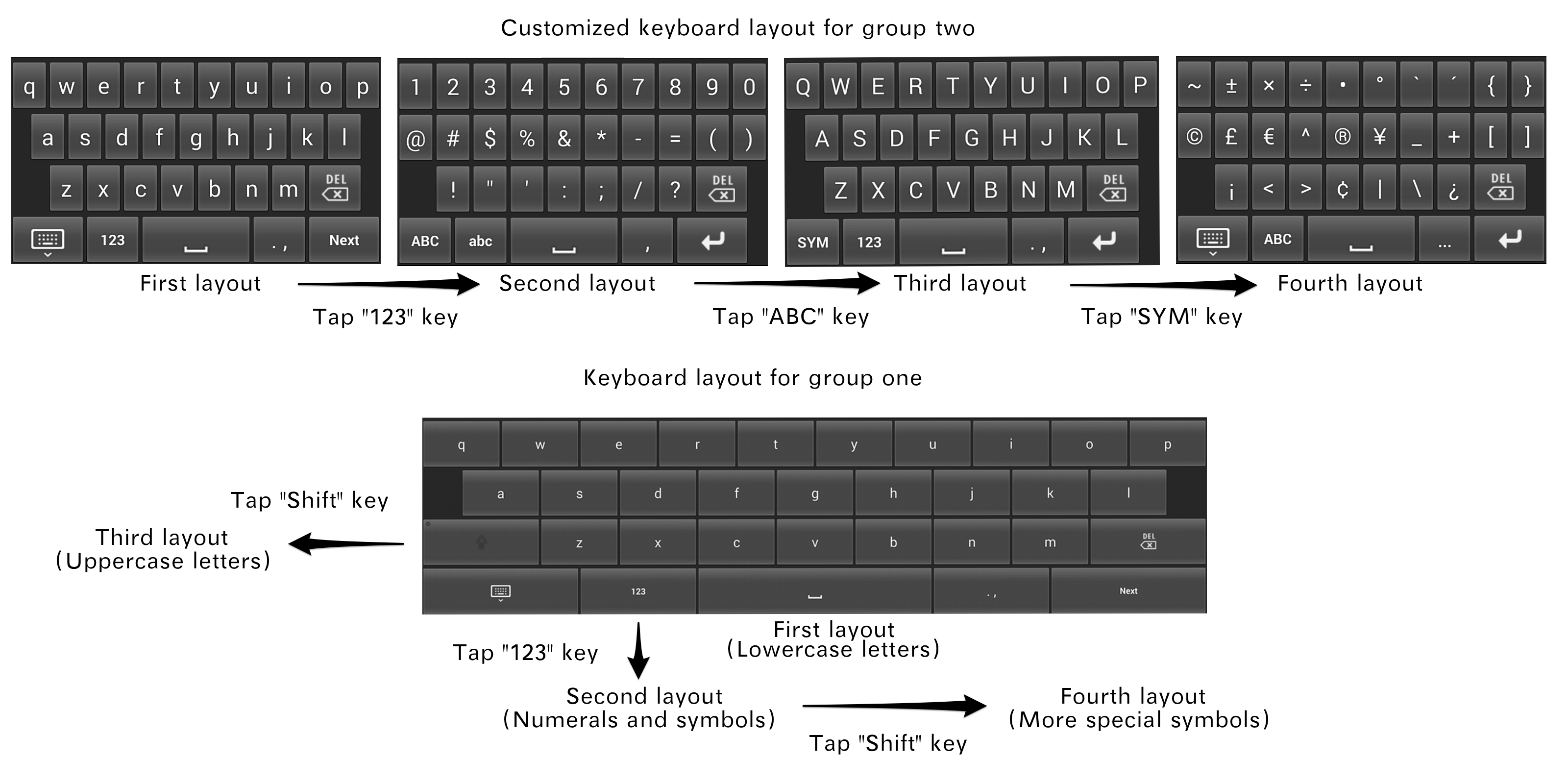}
\caption{\label{fig:customized_kbd_layout} The keyboard layout for the devices in group one and two. Note two groups share the same key positions within each layout, but the structures of the four layouts are different: group one follows the common structure, while group two has a hierarchical structure, that is, to reach the next layout one has to first reach the previous one.}
\vspace{-6pt}\label{fig:customized_kbd_layout}
\end{figure*}

\textit{Software}

We implemented two versions of applications that performed the same task. The application for control group was built with Flask, a Python micro-framework for web development. The application for group one and two was built with Java for Android. Both applications had the same two features: password creation and password recall. In password creation interface, they asked participants to create usernames and passwords for three virtual accounts, including bank, email and online magazine accounts, in the same order. Each virtual account had a different logo, color and short description. In recall interface, both applications asked participants to recall what they created earlier for each account, in a different order, which was determined by Latin square. A ``Give up'' button would show up after 4 failed attempts for each account. Both applications also provided corresponding toast notification for situations such as incorrect password, empty username, successfully recalled, etc.

\subsection{Measurement}

For all groups, we logged all our data into a local file stored on the device. 
We collected username, password text and text entry related data from every participant. Text entry related data included every keystroke and switches between different text entry fields. We also collected the exact timestamp for all entry events. Subjective work load assessment scores were collected using TLX forms.

\subsection{Procedure}

After recruitment, participants were randomly assigned to our three groups separated based on text entry methods. The number of participants for control group, group one and two were 21, 27 and 15, respectively. All studies were conducted in the same office room we setup for this study. Our study consisted two sessions. We explain the procedure of each session as below.

In session one, we asked participants to create a username and a secure password for three different accounts using the text entry methods we provided. 
Specifically, the participants were instructed to create secure passwords, adding that the passwords should be difficult for others to guess while still easy for themselves to remember. They were also informed not to re-use existing passwords. The participants were not given any other instructions about security, and our apparatus did not check any password composition policy. After creating passwords, participants were asked to fill in a TLX form \cite{hart:tlx} regarding the task and perform two distraction tasks we provided, the mental rotation task \cite{mentalrotation}, and a mental countdown. Then, we asked the participants to try to recall their usernames and passwords for the three accounts. The order of the accounts was switched using 3 x 3 Latin square. Then we asked a few survey questions and demographic questions. The detailed procedures for the session one were as follows:

\begin{enumerate}
\item \textbf{Introduction to the Study}. The participants were introduced to the study, which included reading and signing the consent form, discussion of their rights as participants and how they would be compensated.
\item \textbf{Password Creation}. Each participant was given the corresponding text entry method before the session. They were asked to create usernames and passwords for three different virtual accounts: bank, email and online magazine. The order of the accounts was the same for all participants at this step.
\item \textbf{Subjective Workload Assessment} The participants were asked to fill out NASA TLX form.
\item \textbf{Distraction}. The participants were asked to do a mental rotation task and count down from 20 to 0 in mind.
\item \textbf{Password Recall}. Participants were asked to recall usernames and passwords they created in the Password Creation phase above. The order of the accounts were changed with Latin square. Participants were allowed to try as many time as they wanted, and if they cannot remember their usernames or passwords, they could press the ``Give up'' button (showed up after four failed attempts) to proceed to the next account.
\item \textbf{Short Survey}. The participants were asked some questions about how did they create their passwords and other thoughts about the study.
\item \textbf{Demographic questions}. The participants were asked usual demographics questions.
\end{enumerate}

In session two of our study, which was at least 10 days after session one, participants were asked to come back to recall the usernames and passwords they created in session one. The recall procedure was the same as the recall process in session one. After recall process, participants were asked to fill out NASA TLX form and answered a few questions in a short survey.
 
\subsection{Estimating password security}

We estimated the security of our passwords quantitatively with several password security metrics: Shannon entropy~\cite{Shannon:2001:MTC:584091.584093, 394764}, NIST entropy for human-selected passwords~\cite{Burr:2011:SEA:2206278} and scores based on the adaptive password-strength meter (APSM)~\cite{Castelluccia:2012:NDSS:markov}.

Both Shannon entropy and NIST entropy are well-known password strength estimators: entropy has been proposed to be a measure of the randomness of passwords in bits. As a measure of ``uncertainty", Shannon entropy has been used in evaluating security of passwords in cryptographic contexts~\cite{Burr:2011:SEA:2206278}. We used random entropy in our analysis, which was defined as in equation $H = L \times \log_{2} N$, in which $L$ is the length of the password, and $N$ is the possible set of characters.

The NIST entropy is a scheme to evaluate human-selected passwords introduced in NIST Electronic Authentication Guideline~\cite{Burr:2011:SEA:2206278}. The scheme takes into account the fact that passwords were chosen by human beings, who tend to choose passwords that are easily guessed, and even from fairly small dictionaries of a few thousand commonly chosen passwords. According to the NIST guideline we implemented the scheme by assigning different amount of entropy to characters at different positions, each password creation rule contributing a specific amount of entropy and that the entropy of the policy is the sum of the entropy contributed by each rule. In addition, we performed a simple dictionary word check to give the password a bonus entropy if it did not contain checked dictionary word. The dictionary we used here was ``dic-0294", which will be described in later section.

The adaptive password-strength meter (APSM) based on Markov Models is trying to estimate the strength of a password by estimating the probability of the n-grams that compose the password~\cite{Castelluccia:2012:NDSS:markov}. N-gram is a concept in natural language processing: an n-gram is a contiguous sequence of n characters from a given string. Probabilities of n-grams are computed based on a large password dataset, therefore, it introduces certain dependency on the training password dataset. In our implementation, we used the ``rockyou" password dataset to compute the database of probabilities for every n-gram. Also, we chose 4-gram as the element in our database, which was the same as the original paper. The ``rockyou" password dataset, which would be explained in later section, contained over 32 million real passwords. The detailed implementation could be found in the original paper.

There are some other metrics we did not include. Bonneau et al.~\cite{Bonneau:2012:SMI:2437647.2437657} has computed several statistical metrics for password security. However their goal is mainly to estimate the password security in the context of a large password dataset (about 70 millions in the literature). Therefore, their metrics did not fit well in our scenario with a small set of data. The guess-number calculator proposed by Kelley et al.~\cite{Kelley:2012:GAM:2310656.2310715} is trying to predict the number of guesses a cracking algorithm needed to compromise a given password, instead of actually cracking it. Given there lacks of a comprehensive evaluation for the predictor, it is difficult to repeat the work exactly. However, instead of applying the metric directly, we used one of mentioned cracking setup from the work in our actual cracking attacks. The chosen experiment was claimed to simulate an attack with across to a broad variety of publicly available data~\cite{Kelley:2012:GAM:2310656.2310715}. Therefore, we expected it to be a comprehensive attack to reveal vulnerability in our passwords. More detailed description of the setup could be found in the following subsection.

\subsection{Password Cracking attacks}

We performed several actual cracking attacks against our passwords to gain more insights beyond password security metrics. The attacks included: a plain dictionary attack with different types of dictionaries, and two long session offline attacks inspired by previous work~\cite{Weir:2010:TMP:1866307.1866327, Kelley:2012:GAM:2310656.2310715}. As Bonneau et al.~\cite{Bonneau:2012:SGA:2310656.2310721} indicated, the result of cracking attacks also suffered from a lack of comparability and repeatability. Therefore, our goal was to discover both the same and the differences in resistance against attacks for different groups of passwords in general. In our attacks, we used two popular password cracking tools, John the Ripper~\cite{johntheripper} and hashcat~\cite{hashcatsoftware}. We described our collection of dictionaries, and the setup of our attacks as follows. 

\textit{Dictionaries}

We used various dictionaries that are common in previous work. ``dic-0294" was a commonly used English dictionary~\cite{outpost9}. ``all" and ``mangled" dictionaries were free and paid dictionaries from openwall website\footnotemark. \footnotetext{http://www.openwall.com/wordlists/} ``all" contained words chosen by openwall from 21 different languages plus a list of frequently used passwords, and ``mangled" was a hand-tuned wordlist containing nearly 4 million password candidates with different mangle rules applied to various dictionaries from openwall. ``rockyou" dictionary included about 32 million passwords leaked from the game website RockYou. ``facebook" was a full list of names of search-able user from the social network website Facebook~\cite{facebookindex2010}. For each name ``facebook" had four formats: first name, last name, the first name plus last initial, and first initial plus last name. ``myspace" contained passwords from a phishing attack against MySpace website. ``inflection"\footnotemark \footnotetext{http://wordlist.sourceforge.net} was a list of words along with their different grammatical forms such as plurals and past tense. 

\textit{Dictionary attack}

First, we applied a plain dictionary attack with different combinations of our dictionaries. A plain dictionary attack involves simply comparing each word/password from the dictionary to the target password in hashed format and see if they are the same. The performance of a plain dictionary attack relies heavily on the relevancy of the dictionary and the target passwords. We classified mentioned dictionaries into three types. The first attack with ``Words", which contained common words from different languages, aimed at easy passwords which only had single dictionary word; the second with ``Facebook", contained the entire directories from the website, aimed at passwords made with actual names of people or organizations, and popular phrases; the third attack with ``Passwords", which contained common passwords and real leaked passwords, aimed at compromising common and naive passwords.

\textit{Long session offline attack}

We also applied two long session offline attacks on our passwords.

The first attack involved generating guesses based on an modified ``Single mode" mangle rules, which originally from John the Ripper, with the ``dic-0294" dictionary as input. The ``Single mode" rules set contained a set of rules to modify words including login names and home directory to generate a large size of guesses~\cite{johntheripper}. The modified version, made by Weir~\cite{jtrmodifiedsingle2010}, was said to be optimized for English dictionary. We followed the experiment setup in~\cite{Weir:2010:TMP:1866307.1866327}. 

The second attack applied an advanced password crack algorithm: the state-of-art probability password crack tool developed by Weir et al.~\cite{Weir:2009:PCU:1607723.1608146, Weir:2010:TMP:1866307.1866327}. The tool generated password guesses in the order determined by various rules in structures, characters, digits and symbols.
The tool needed to be trained in advance for developing the model of probability, then one could apply the model to a given dictionary to generate guesses. We used a similar model from experiment P4 conducted by Kelley et al.~\cite{Kelley:2012:GAM:2310656.2310715}. Note though while the experiment from the original work was an estimation of number of guesses against each target password, we actually generated those guesses against our passwords. 

The detailed setup of our second attack is as follows: 

\begin{enumerate}
\item using passwords with length longer than 8 from ``myspace", ``all" and ``mangled" dictionaries to train the probabilities of character-type structures;
\item using ``myspace", ``all", ``mangled" and ``rockyou" dictionaries to train the probabilities of digits and symbols in passwords;
\item finally using ``myspace", ``dic-0294", ``inflection", ``rockyou", ``all" and ``mangled" dictionaries as input strings.
\end{enumerate}

The number of guesses generated in two attacks were 1000M and 20000M respectively. We chose so to simulate one quick attack with easy-to-find resource while another longer attack with optimal strategies and more resources.

\subsection{Statistical models}

Here we described the assumptions and statistical models we used in our analysis. We assumed the data we collected from each text entry method group came from normal distribution, which enabled us to apply classic statistical models. The reason we held this assumption because all our groups had a large enough sample size ($>30$ per group), indicating they were representative for each population. In cases where required assumptions could not be met, we used corresponding robust methods. Note our hypothesis was on the effect of text entry methods, therefore our main focus was the text entry method variable.

For analysis of categorical dependent variables, we used Pearson's chi-square test. The Pearson's chi-square test is used to examine the null hypothesis that the distribution of frequencies one observed in certain samples is consistent with the targeted distribution~\cite{doi:10.1080/14786440009463897}. 
Pearson's chi-square test assumes minimum expected frequency in the testing categorical data is larger than 5.
There were cases where this assumption could not be held. In such cases we used Fisher's exact test~\cite{doi:10.2307/2340521}, which is the robust equivalent of chi-square test. Since given categorical tests does not compatible with repeated measure variables~\cite{isbn:9781446200469}, we did not include account type variable in such tests.

For analysis of continuous dependent variables, we applied chi-square tests on multilevel models. Multilevel model (or mixed model) is a generalization of linear regression model. It compares the effect of several variables on the group mean of data, allowing data from participants to be organized at more than one level~\cite{barbarataba2007}, or one variable nested within another, which fits in our experiment setup. For each dependent variable, we constructed different multilevel models by adding one variable at a time. We also constructed a model with interaction effect of our two variables. Then we compared the fit of these models to our data using chi-square fit test. If the result showed significant difference between two models, which should differ by whether having one particular variable or not, then we concluded that variable had significant effect on the dependent variable. Such models also enabled us to look at pair-wise comparisons within one model to gain more insights, given we have three levels in both two variables.

For analysis of several correlated dependent variables, we applied the multivariate analysis of variance (MANOVA). MANOVA is a generalized form of univariate analysis of variance (ANOVA) and is used when there are two or more dependent variables~\cite{Stevens:1986:AMS:21800}. In our analysis, we used the MANOVA with the Pillai-Bartlett trace as the statistic~\cite{isbn:9781446200469}.
Note for all significance tests performed, we chose $p=0.05$ as the threshold for a statistically significant result.

\let\oldtabular\tabular
\renewcommand{\tabular}{\small\oldtabular}

\section{Results}

In total we have collected 189 passwords from 63 participants. In this section we begin with a basic analysis on password structures and probability of individual key appeared in a password, then proceed to the analysis using standard and recent password strength metrics, the result of our password cracking attacks, and the analysis of TLX data. Throughout the section, we mostly focus on the effect of different text entry methods to password security.

\subsection{Structures}

Table~\ref{tab:basics} shows the length of the generated passwords and also the amount of characters of different types per password across groups. The result demonstrates a notable difference in password length and amount of lowercase characters between group two and other two groups.

\begin{table}[tbph]
\begin{center}
\scalebox{0.9}{
  \begin{tabular}{ p{25mm} | p{15mm} | p{15mm} | p{15mm}}
    \hline
    Metric & Control group & Group one & Group two \\ \hline
    Length (mean) & 10.13 & 10.05 & 12.27 \\ \hline
    Lowercase letters & 6.19 & 6.37 & 8.40 \\ \hline
    Digits & 2.59 & 2.31 & 1.67 \\ \hline    
    Uppercase letters & 0.75 & 0.78 & 1.22 \\ \hline
    Special symbols & 0.32 & 0.30 & 0.49 \\ 
    \hline
  \end{tabular}
  }
  \caption{The mean of password length, amount of characters of different types per password across groups.}
    \label{tab:basics}
\end{center}
\end{table}

We fitted different multilevel models take into account different variables, and then tested if any variable had significant effect to the fit of the model by conducting chi-square test. Our variables were text entry method (3 levels) and account type (3 levels). The same set of models was applied to listed dependent variables separately. The result is shown in Table~\ref{tab:multilevel_element}. It shows only account type variable has a significant main effect on the number of digit appeared in single password across groups ($\chi^{2}(2)=10.98$, $p=0.0041$). The text entry method variable has an effect very close to significant on the number of lowercase letters across groups ($\chi^{2}(2)=5.92$, $p=0.052$). 

\begin{table}[tbph]
\begin{center}
\scalebox{0.9}{
  \begin{tabular}{ l | c | c | c }
    \hline
    Metric & Variable & $\chi^{2}(2)$ & $p$ \\ \hline
    \multirow{3}{*}{Length}
    & text entry method type & 1.47 & 0.49\\
    & account type & 2.49 & 0.29\\
    & interaction & 4.12 & 0.39\\ \hline
    \multirow{3}{*}{Lowercase letters}
    & text entry method type & 5.92 & 0.052* \\
    & account type & 5.14 & 0.08\\ 
    & interaction & 5.55 & 0.24\\ \hline
    \multirow{3}{*}{Digit}
    & text entry method type & 2.75 & 0.25\\ 
    & account type & 10.98 & 0.0041*\\ 
    & interaction & 3.97 & 0.41\\ \hline
    \multirow{3}{*}{Uppercase}
    & text entry method type & 0.89 & 0.64\\ 
    & account type & 2.08 & 0.35\\ 
    & interaction & 7.20 & 0.13\\ \hline
    \multirow{3}{*}{Special symbol}
    & text entry method type & 0.24 & 0.89\\ 
    & account type & 5.19 & 0.07\\ 
    & interaction & 3.58 & 0.47\\
    \hline
  \end{tabular}
  }
  \caption{Result of chi-square fit test on multilevel models for amount of characters of each type per password. Variables in the model were text entry method type and account type. The term ``interaction" stands for the interaction effect between the two factors.}
    \label{tab:multilevel_element}
\end{center}
\end{table}

We then further examined the effect of text entry methods on amount of lowercase letters by setting up planned contrasts. The contrasts looked for differences in the amount of lowercase letters of a password between control group and group one, $b=0.43, t(60)=0.45, p=0.65, r=0.058$, and control group and group two, $b=3.35, t(60)=2.99, p=0.004, r=0.36$. This result shows a significant effect of our variable on the amount of lowercase letters between control group and group two. Specifically, passwords created from group two had significantly more lowercase letters than that from control group.

Next, we examine the categories of passwords each group generated. We defined the category of a password by types of characters it contained. The category of a password reveals the complexity in its structures: a password containing only one type of characters has a much simpler structure than one with several different types of characters. Table~\ref{tab:categories} summarizes our definition of categories.

\begin{table}
\begin{center}
\scalebox{0.9}{
  \begin{tabular}{p{30mm} | p{50mm}}
    \hline
    Category & Description\\ \hline
    loweralphanum & only contains lowercase letters and digits\\ \hline
    loweralpha & only contains lowercase letters\\ \hline
    mixedalphanum & contains lowercase and uppercase letters and digits\\ \hline
    loweralphaspecialnum & contains lowercase letters, special symbols and digits\\ \hline
    all & contains lowercase and uppercase letters, special symbols and digits\\ \hline
    mixedalpha & only contains lowercase and uppercase letters\\ \hline
    others & types other than mentioned ones\\
    \hline
  \end{tabular}
  }
  \caption{Definition of each category of passwords. All types with very low occurrence in our passwords were aggregated into ``others" category.}
    \label{tab:categories}
\end{center}
\end{table}

\begin{figure}[!Htbp]
\includegraphics[width=0.9\linewidth]{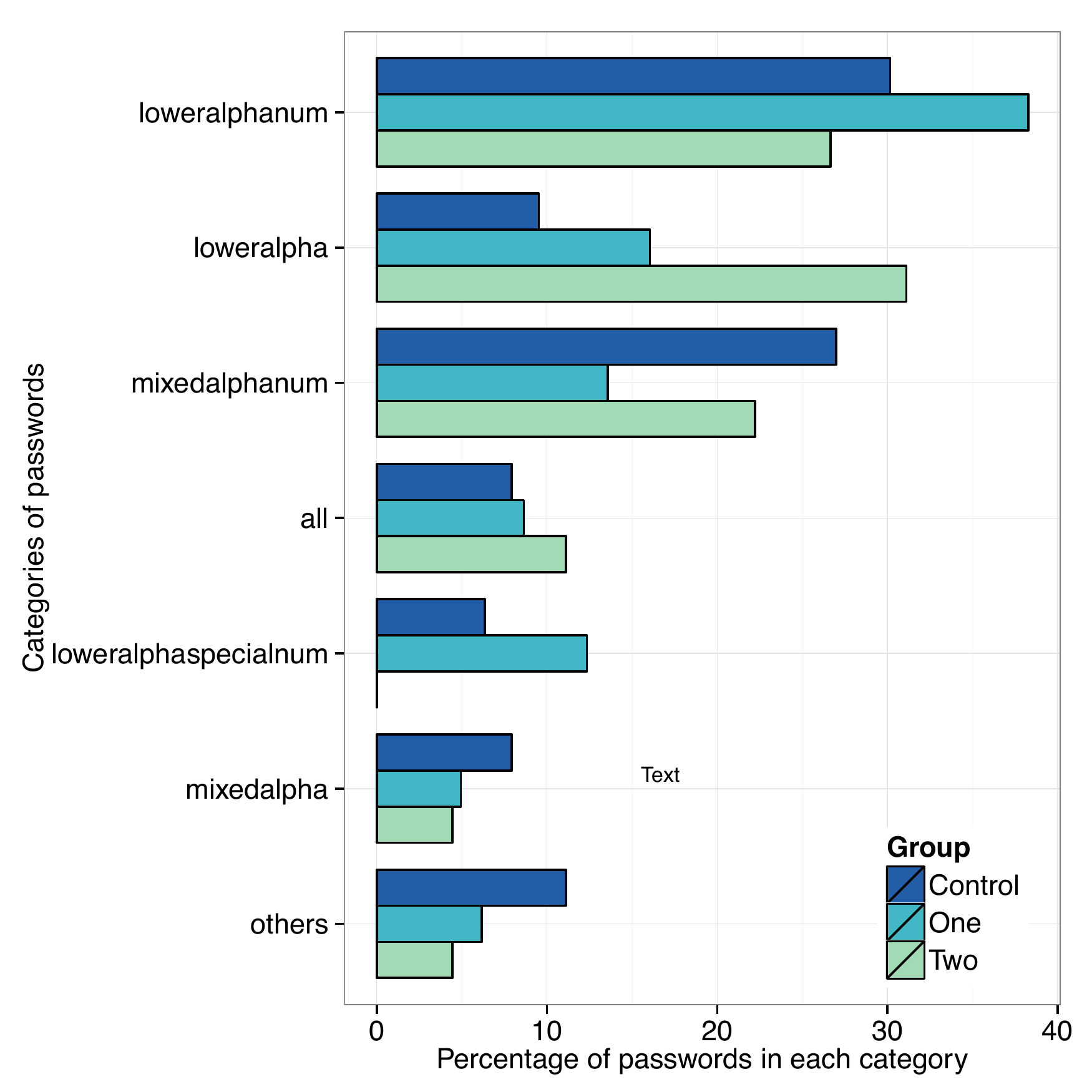}
\caption{\label{fig:bar_categories} A comparison of distribution of passwords in different categories for each group.}
\vspace{-9pt}
\end{figure}

Figure~\ref{fig:bar_categories} demonstrates the distribution of passwords in defined categories across groups of passwords. According to the figure, the most common category for passwords from both control group (30.2\%) and group one (38.2\%) are passwords contain only lowercase letters and digits (\emph{loweralphanum}), while that for group two is passwords contain only lowercase letters (\emph{loweralpha}) (31.1\%). In addition, while other two groups both have a certain amount of passwords contain lowercase letters, special symbols and digits (\emph{loweralphspecialnum}), group two does not have passwords in that category at all.

To quantify the difference of the overall distribution across groups, we applied Fisher's exact test with text entry method as variable, given our data did not meet the minimum expected frequency. The result, ($p=0.048$, $FET$), shows a significant effect of text entry method over the frequency distribution of our passwords in mentioned categories.

\subsection{Probability of keys appeared in passwords}

\begin{figure*}[htbp]
\begin{center}
\includegraphics[width=0.7\linewidth]{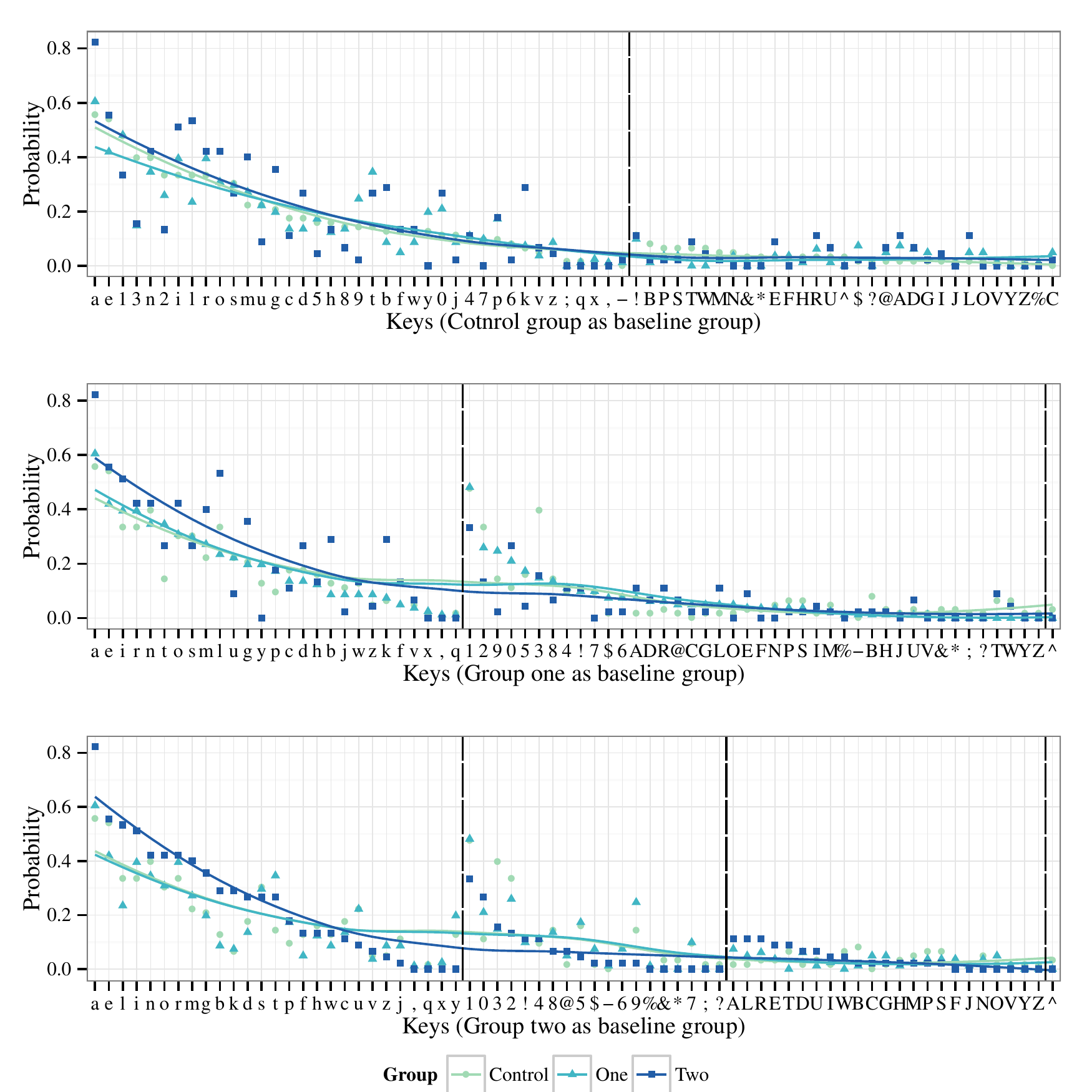}
\caption{\label{fig:probability_v_rank_all} The probabilities of keys across groups. Three graphs display the same set of data but use different group as \emph{baseline group}: control group (top), group one (middle), group two (bottom). In each graph, probability of each key across groups are plotted as dots. Vertical lines indicate the order of layouts (first, second, etc) based on that of \emph{baseline group}. Within each layout, the order of keys on x axis is ranked by probability of each key from the corresponding \emph{baseline group} Curve lines in each graph represents a non-linear fit to data of each group.} 
\vspace{-9pt}
\end{center}
\end{figure*}

Previous subsection indicates a significant difference in the amount of lowercase keys across groups. To examine more detailed difference among keys in our passwords, we analyzed the probability of each key appeared in a password for our data. The probability of a key within one group is defined as the number of passwords contained the key over the total number of passwords created by that group. We also included the comparison between keys with different reachability. As mentioned previously, reachability of a key is defined as how many key presses one needs to reach that key from the first layout. For keys from a keyboard, the value of reachability depends on which layout they are in. For example in our designs, the same key ``A", is in the second layout in control group and group one, but in the third layout in group two. Figure~\ref{fig:probability_v_rank_all} demonstrates the result. 

The figure shows that when we use control group as \emph{baseline group}, the nonlinear fitted lines of three groups do not differ much. However we find that probabilities of the same across groups are quite different. For most of keys, its probabilities of group one and group two ``oscillate" around that of control group. Such ``oscillation" canceled low probabilities and high probabilities with each other for the experimental groups in some degree. To compare, when we use experimental groups as \emph{baseline group}, the difference in the fitted lines becomes visible, especially with group two as \emph{baseline group}: in both the middle and bottom graph, the probabilities of lowercase keys of group two are more ``skewed" than that of other two groups; also, the probabilities of digits and most symbol keys of group two are visibly lower than that of other two groups. This suggests people from group two created passwords had a preference over a small subset of lowercase keys compared with other groups. We explained this result in more detail later in discussion.

\subsection{Password security by quantitative metrics}

\begin{figure}[htbp]
\includegraphics[width=0.9\linewidth]{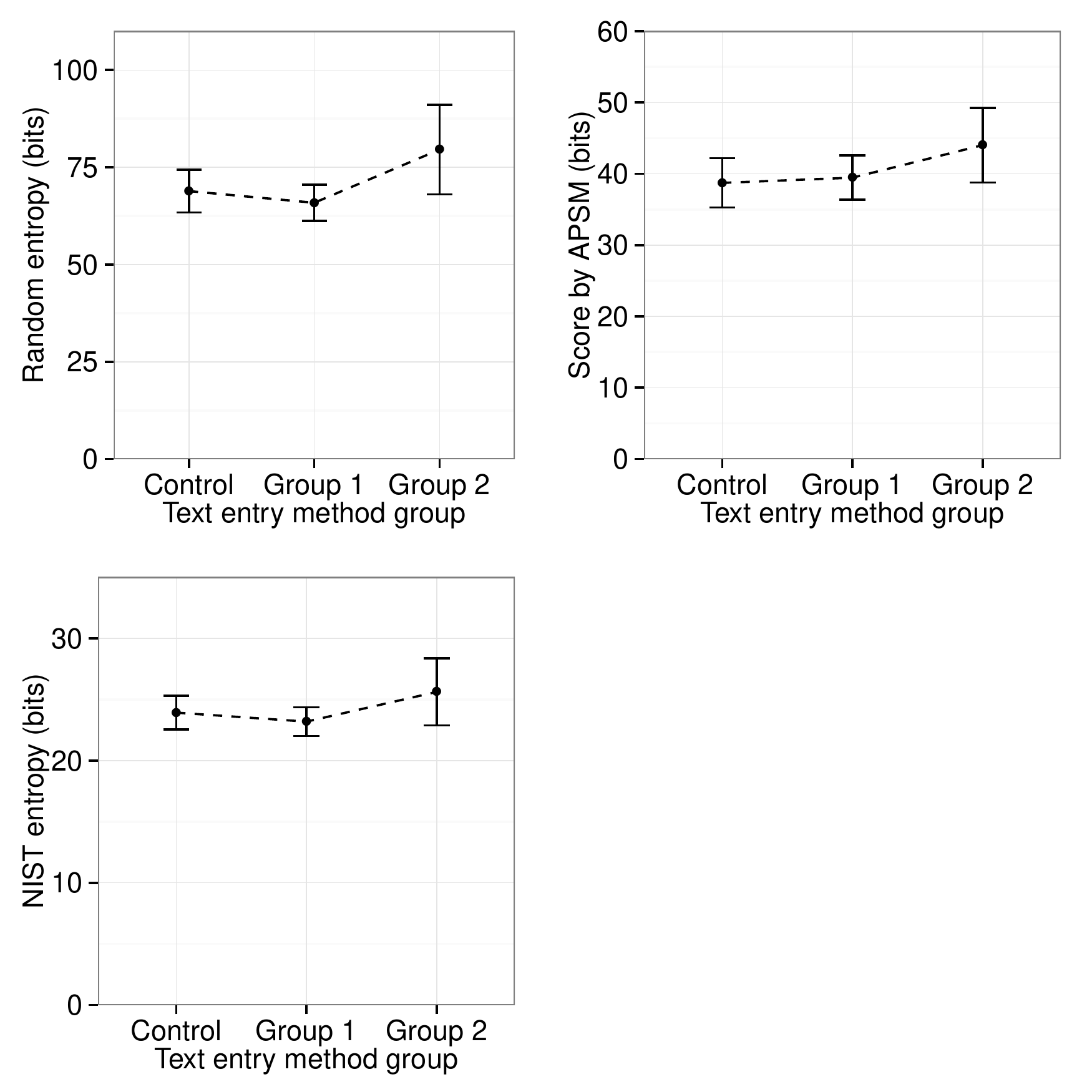}
\caption{\label{fig:strength_length} The mean score of three password security metrics across groups. They are: random entropy (upper left), NIST entropy (bottom left) and score from Adaptive Password-strength meter (APSM) (upper right). The error bar indicates 95\% confidence interval.}
\vspace{-9pt}
\end{figure}

Next, we analyzed our passwords with two standard and a recent password security metrics. For each password, we computed random entropy, NIST entropy and score computed using adaptive password-strength mater based on Markov model (APSM). The mean scores and corresponding confidence intervals of the result are shown in Figure~\ref{fig:strength_length}. According to the graph, scores of passwords of group two are consistently higher than that of other two groups. However, most of means stay within the confidence interval of the value of other groups, indicating the differences among groups are limited. 

To quantify the difference, we fitted different multilevel models take into account different variables, and then tested if any variable had significant effect to the fit of the model by conducting chi-square test. Our variables were text entry method (3 levels) and account type (3 levels). The same set of models was applied to listed dependent variables separately. We present the result in Table~\ref{tab:multilevel_4}. It shows that the text entry method type did not have significant main or interaction effect on any of our metrics. The account type variable, on the other hand, has significant main effect on NIST entropy ($\chi^{2}(2)=7.12$, $p=0.0284$) and the scores from APSM ($\chi^{2}(2)=8,49$, $p=0.0143$). However, the pair-wise effect size of account type variable on both scores from APSM ($r_{bank, email}=0.07$, $r_{bank, magazine}=0.16$) and NIST entropy ($r_{bank, email}=0.04$, $r_{bank, magazine}=0.1$) were quite small. Therefore, in the scope of our data and these password metrics, different groups did not demonstrate significant difference. 
However, we threw reasonable doubt on the validity and comprehensiveness of those metrics as password security measurement, which was proved by the result of our cracking attacks. Detailed explanation could be found in discussion section.

\begin{table}[thbp]
\begin{center}
\scalebox{0.9}{
  \begin{tabular}{ l | c | c | c}
    \hline
    Metric & Variable & $\chi^{2}(2)$ & $p$\\ \hline
    \multirow{3}{*}{Random entropy}
    & text entry method type & 1.88 & 0.39\\
    & account type & 2.20 & 0.33\\ 
    & interaction & 4.09 & 0.39\\ 
    \hline
    \multirow{3}{*}{NIST entropy}
    & text entry method type & 1.19 & 0.55\\ 
    & account type & 7.12 & 0.03*\\ 
    & interaction & 1.26 & 0.87\\ 
    \hline    
    \multirow{3}{*}{Scores of APSM}
    & text entry method type & 1.33 & 0.51\\
    & account type & 8.49 & 0.01*\\ 
    & interaction & 2.78 & 0.60\\ 
    \hline
  \end{tabular}
  }
  \caption{Result of chi-square fit test on multilevel models applied on three security metrics. Variables in the model were text entry method type and account type. The term ``interaction" stands for the interaction effect between terminal type and account type.}
    \label{tab:multilevel_4}
\end{center}
\end{table}

\subsection{Cracking attacks}

In this section, we present the results of comprehensive cracking attacks performed against our password dataset. Table~\ref{tab:dictionary_attack} shows the result of plain dictionary attacks. The performance of ``Words" and ``Facebook" attack was very limited across all groups, except ``Facebook" attack on passwords from group two. The ``Password" attack worked much better compared with first two attacks against control group and group one, while had very limited improvement against group two.

\begin{table*}[tbph]
\begin{center}
  \begin{tabular}{ l | c | c | c | c | c }
    \hline
    Name & Include & Size & Control group (63) & Group one (81) & Group two (45) \\ \hline
    Words & ``dic-0294", ``all", ``inflection" & 4.1M & 4 (6.3\%) & 4 (4.9\%) & 4 (8.9\%) \\
    Facebook & ``facebook" & 37.3M & 3 (4.8\%) & 6 (7.4\%) & 7 (15.6\%) \\
    Passwords & ``mangled", ``rockyou" & 54.8M & 15 (23.8\%) & 12 (14.8\%) & 8 (17.8\%) \\
    \hline
  \end{tabular}
  \caption{Results of plain dictionary attack with different dictionaries. ``Include" included all dictionaries we used in each attack. The size is the number of unique entries each combined dictionary has.}
    \label{tab:dictionary_attack}
\end{center}
\end{table*}

\begin{figure}
\includegraphics[width=0.9\linewidth]{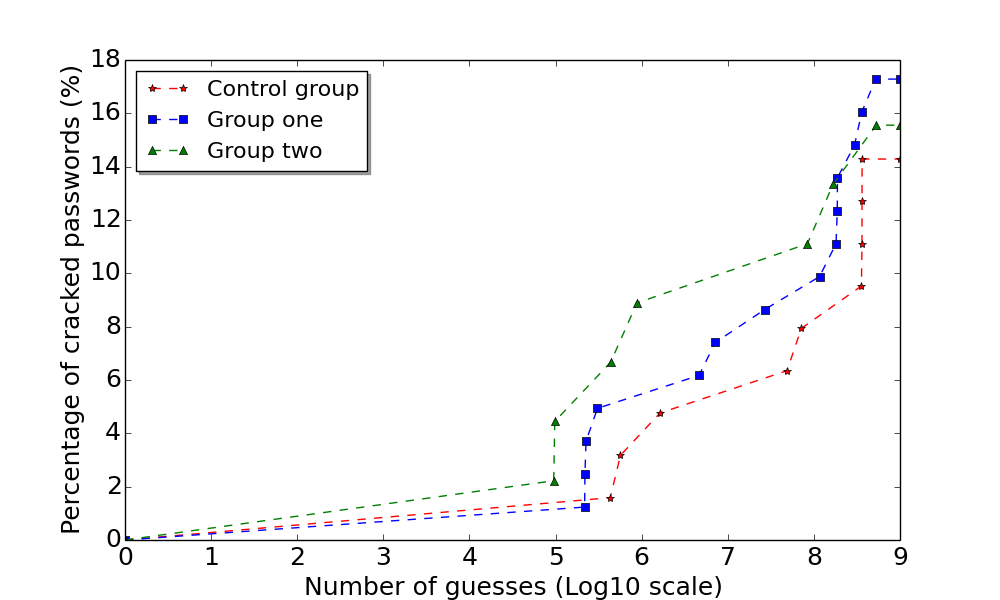}
\caption{\label{fig:msingle_percentage} The percentage of passwords cracked by our first offline attack. The x axis is in in log scale.
The final percentage of cracked passwords for control group, group one and group two are 14.2\%, 17.3\% and 15.6\%, respectively.}
\vspace{-9pt}
\end{figure}

\begin{figure}
\includegraphics[width=0.9\linewidth]{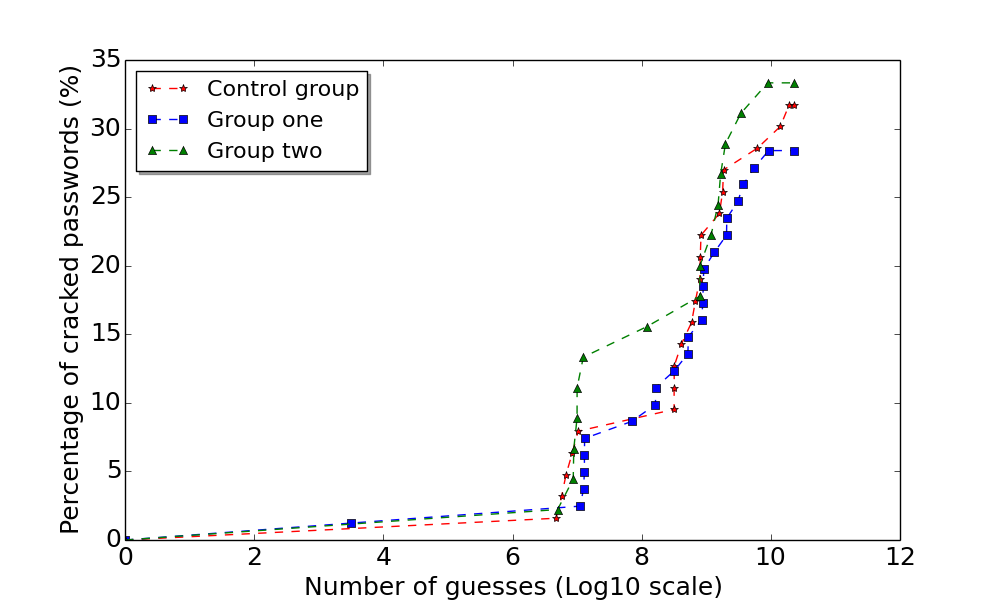}
\caption{\label{fig:guess_again} The percentage of passwords cracked by Weir et al.'s algorithm vs. the number of guess, per group. The final percentage of cracked passwords for control group, group one and group two are 31.7\%, 33.3\% and 28.4\%, respectively.}
\vspace{-9pt}
\end{figure}

The results of two long session offline attacks are shown in the Figure~\ref{fig:msingle_percentage} and Figure~\ref{fig:guess_again}, respectively. According to the figures, although the lowerbounds of resistance (the number of guesses of the first cracked password) are different, the percentages of cracked passwords across groups show similar patterns compared with each other. One exception is a distinctive spike of the percentage of cracked passwords in the second attack against group two between the range of 1M ($10^{6}$) to 100M ($10^{8}$) in number of guesses.

\begin{figure}[tbph]
\includegraphics[width=0.9\linewidth]{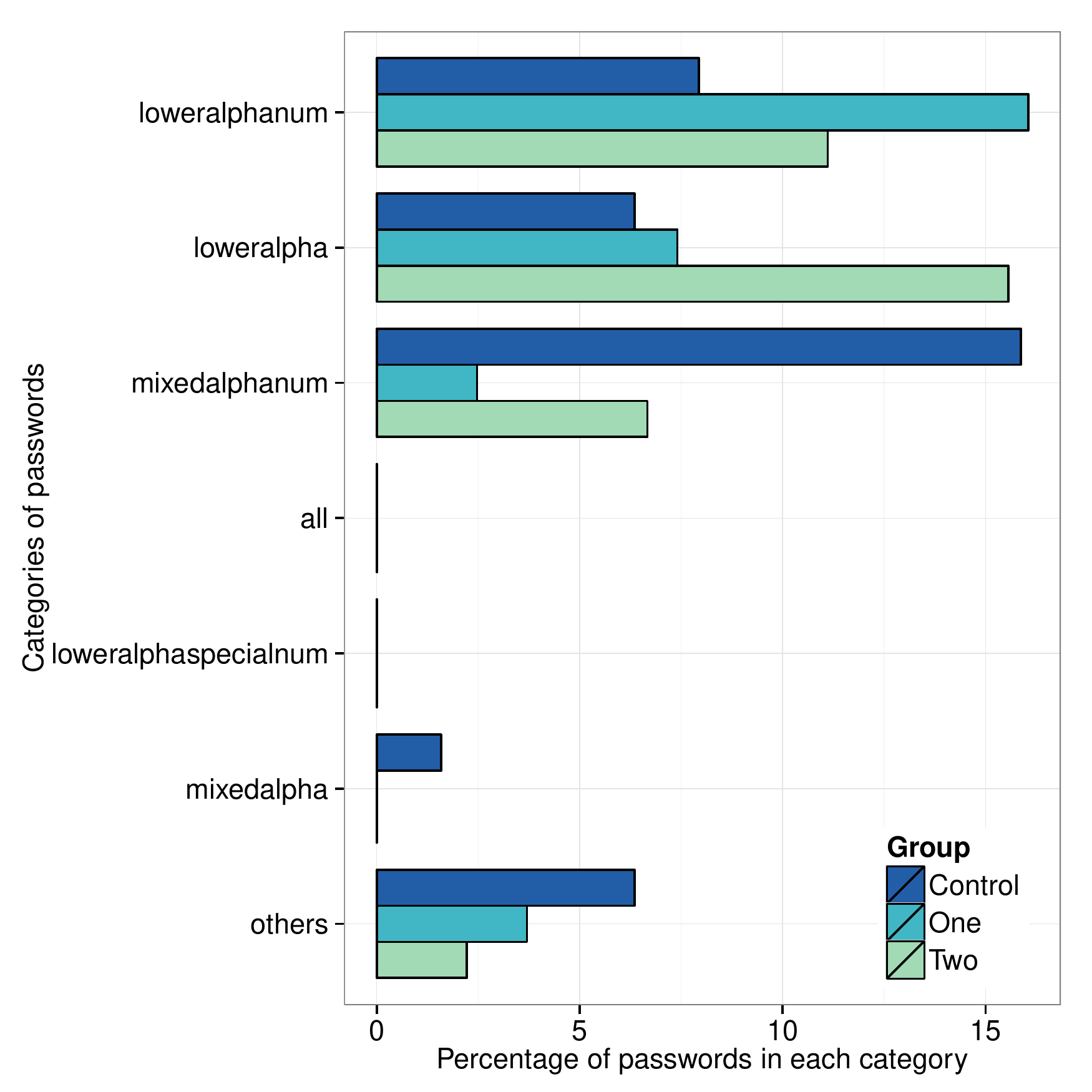}
\caption{\label{fig:bar_categories_cracked} A comparison of percentages of cracked passwords in different categories across groups. Cracked passwords here included passwords compromised from all our attacks.}
\vspace{-9pt}
\end{figure}

When we combine cracked passwords from all attacks together, the total number of cracked passwords for control group, group one and group two are 24 (38.1\%), 24 (29.6\%) and 16 (35.6\%). Figure~\ref{fig:bar_categories_cracked} shows the distribution of all cracked passwords into different categories across groups, in which we see quite different percentages in different categories across groups. Particularly, the category with the largest percentage of cracked passwords is different for all three groups: \emph{mixedalphanum} (passwords contain uppercase letters, lowercase letters and digits) (10, 15.9\%), \emph{loweralphanum} (13, 16.0\%) and \emph{loweralpha} (7, 15.6\%) for control group, group one and two, respectively.

To see the result of whether or not a password would be cracked by our cracking attacks was different across groups based on our data, we applied chi-square test on the number of cracked passwords across groups. The result shows it is likely that the types of text entry method do not have any significant effect on whether or not a password would be cracked by our attacks, ($\chi^{2}(2)=1.2$, $p=0.55$).

In addition, we applied categorical test to see if there exists any significant effect of text entry method variable on the distribution of cracked passwords in different categories across groups. We applied Fisher's exact test, given that the data did not meet the minimum expected frequency. The result shows that a significant effect exists ($p=0.031$, FET). Therefore, it is highly likely that the same set of attacks we performed has different effect on passwords created using different text entry methods.

\textit{Task load}

\begin{figure}
\includegraphics[width=\linewidth]{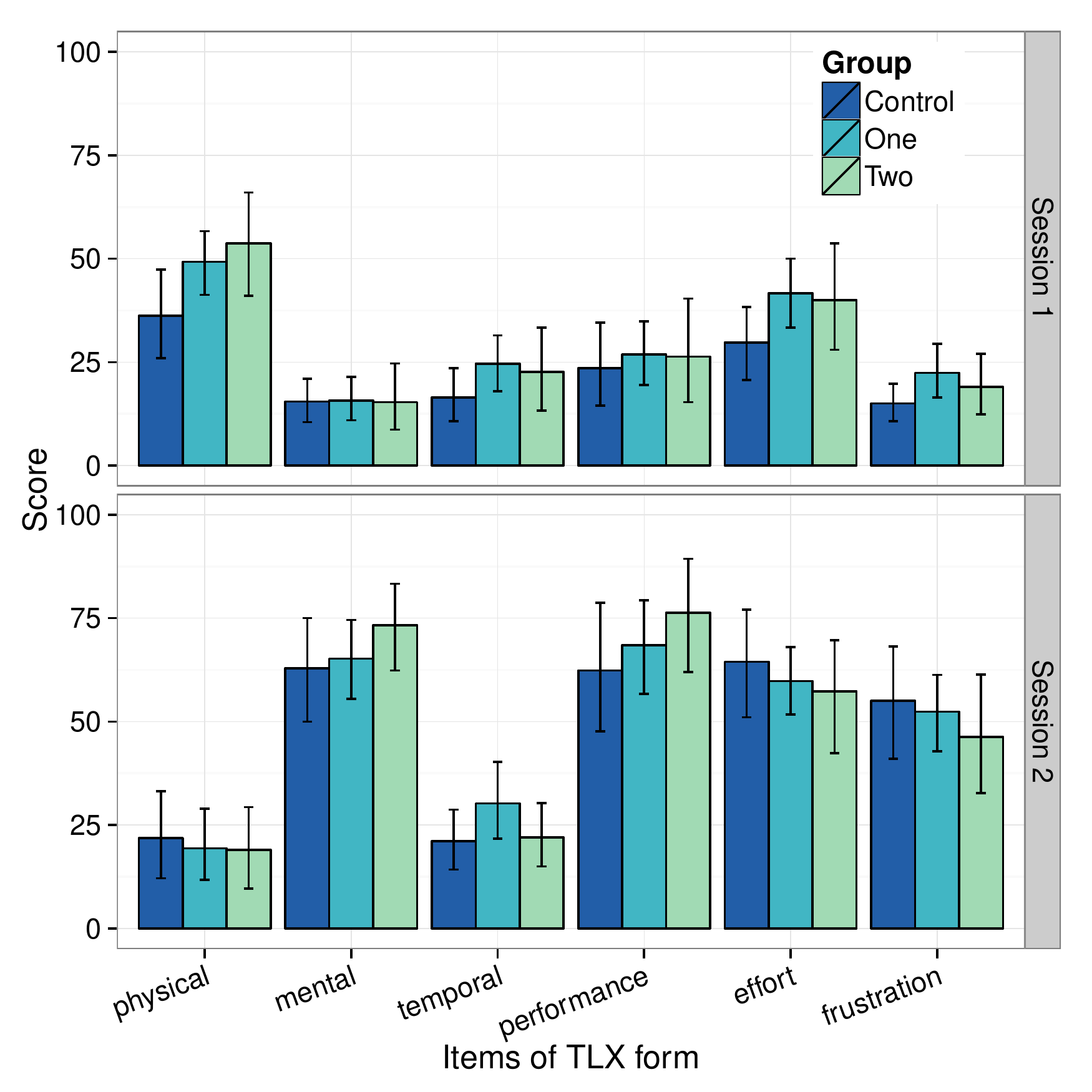}
\caption{\label{fig:tlx_by_item_by_terminal_by_session} Mean score for each item in TLX form in session one and two. The error bar stands for 95\% confidence intervals based on a bootstrap (i.e. not assuming normality).}
\vspace{-9pt}
\end{figure}

In each session we asked participants to fill out the TLX form. We use TLX forms to evaluate the complexity of tasks we designed.  These questions revealed participants' subjective assessment towards tasks in the study, which we summarize below. Figure~\ref{fig:tlx_by_item_by_terminal_by_session} shows the mean scores for each question of TLX form for both session one and two. 

Given items in TLX form were correlated, we applied MANOVA test with the text entry method as variable on the six items together, for session one and two, respectively. The result indicated a non-significant effect of text entry method type on the scores of TLX assessment both for session one, $V=0.21, F(8, 116)=1.70, p=0.11$, and session two, $V=0.28, F(12, 100)=1.37, p=0.19$. Therefore, we concluded that it was highly likely participants in groups did not feel different about the subjective task load of the experiment they participated in. This demonstrates our choice of text entry methods provided similar usability across groups.

\section{Discussion and Conclusions}

Our main findings show that text entry methods affect the security of passwords.

\textbf{Text Entry Method Affects Password Security.}  The structures of passwords created using different text entry methods showed statistically significant differences. This was revealed by the distribution of passwords in different categories across groups $p=0.048$, FET). According to Figure~\ref{fig:bar_categories}, more than half of passwords created from group one ($\sim$54\%) and two ($\sim$57\%) were in categories \emph{loweralphanum} and \emph{loweralpha}. This in contrast to the control group, which had only $39\%$ in the same categories. In particular, when we looked at the percentages of passwords in \emph{loweralpha}, the simplest category, we found that $31\%$ of passwords in group two was in this category. The control group had only $9.5\%$ in the same category. This is emphasized by the significant effect ($p=0.04$) of text entry method variable had on the amount of lowercase letters per password between the control group and group two.

Our analysis of structures was followed by the analysis of individual keys present in passwords. Figure~\ref{fig:probability_v_rank_all} shows the different probabilities of each key in our passwords across groups. The most important finding from this figure is how probabilities of keys become more skewed and squeezed as the difficulty of reaching keys in the text entry method increases. The difference is most visible in the third graph of the figure, when the group two is the \emph{baseline group}, as shown by the fitted lines of groups. In the graph, most of keys before ``w" have a higher probability appearing in passwords from group two, while ones after ``w" have lower probability compared with other two groups. This suggests that people from group two tended to focus on a small subset of lowercase keys when they composed their passwords. In contrast, participants in the other groups chose a broader range of lowercase keys. This also holds for uppercase keys: some uppercase keys had higher probabilities to appear in group two, while other uppercase keys had low or even zero probability. In addition, most digit and symbol keys had low or zero probability of appearing in passwords generated by group two.

We also found that the passwords showed different resistance against cracking attacks. Recall, we categorized the generated passwords to several categories, including ``only contains lowercase letters and digits'', ``contains lowercase and uppercase letters and digits'', etc. Within these categories, the distribution of the cracked passwords was significantly different ($p=0.031$, FET). In particular, the category with the largest percentage of cracked passwords was distinct for all three groups. Since our passwords represent the population which we drew samples from, the same type of vulnerability exists in the population, too. This means that if attackers would know the text entry method used to create the passwords, the corresponding vulnerability could be easily exploited by their attacks. As an example, see Figure~\ref{fig:guess_again}. When the number of guesses is between 1M ($10^{6}$) to 100M ($10^{8}$), there is an unusual spike that exists only in group two's cracked passwords. By examining cracked passwords, we find this is precisely the vulnerability of group two: all cracked passwords in the range are in the \emph{loweralpha} category. We already showed that group two's passwords had significantly more lowercase letters, and participants in this group focused on a small set of lowercase keys when they composed their passwords.

\textbf{Entropy and Markov Model Based Metrics.} There is a contradiction when we compare the results from computing quantitative security metrics to the cracking attacks. In particular, the result of statistical tests on all three quantitative password security metrics were non-significant, yet by issuing cracking attacks we found considerable differences in resistance against cracking attacks across groups. Such difference mostly due to the fact that most quantitative metrics estimate password security by a single value. Using such metrics, two very similar passwords could result in quite different ``score" in security. For example, ``vowelword" and ``bonsjones" were both lowercase-only letters consisted of two English words; however, using APSM, their scores are 50 bits and 30 bits, respectively. Considering the mean score of APSM for our passwords overall are only 40 bits, a difference of 20 bits would be very significant. Yet, one could see little difference between passwords themselves. Therefore, values computed using quantitative metrics could be misleading and should be carefully interpreted, or accompanied with alternative analysis such as analysis of password cracking attacks. Our findings confirm previous recent studies~\cite{Weir:2010:TMP:1866307.1866327, Bonneau:2012:SGA:2310656.2310721} that question using entropy-based metrics to evaluate password security. Our study also indicates that APSM might need to be also reconsidered in particular for small-scale password datasets.

\textbf{Guidelines for Designing Text Entry Methods for Password Creation.} Based on our results, we can introduce guidelines for the design of text entry methods. The fact that experimental groups utilized characters other than lowercase letters less than the control group indicates that the easier access to other characters could be important for users to generate more diverse passwords. Therefore, a simple design modification could be including digits or some special symbols in the first layout of the keyboard, in order to encourage people to choose them over lowercase letters. This is especially important for keyboards on platforms with small input areas, such as smartphones, as our study showed the large amount of \emph{loweralpha} passwords created by them.

However, more diverse passwords do not directly equal to better security, as indicated by our password cracking results. Many diverse passwords generated in the control group were passwords consisting of simple words modified with simple mangling rules,. These passwords were vulnerable to cracking attacks. Therefore, automatically detecting naive mangling rules would be desirable when generating password. We believe such system is quite feasible practically if it only focuses a few very simple mangling rules, for example, capitalized first letters or appending of a single digit.

\textbf{Limitations.} A recent study by Fahl et al.~\cite{Fahl:2013:EVP:2501604.2501617} compared real passwords to those generated in an experiment, finding that about 30\% of their subjects did not behave as they do in real life. However, the authors concluded that laboratory studies generally create useful data. We used a random assignment of subjects to the three terminal conditions, the contribution of such users to the dataset should be even. Our study also included a recall session after at least 10 days had passed for encouraging participants to create realistic passwords.

\textbf{Conclusions.} 
We have presented a comprehensive analysis of passwords created with different text entry methods.
Although we have shown that an effect of text entry methods does exist, more work is needed to understand how and why the different features of the text entry methods contribute to password security. Is it possible to design a method for password entry that specifically encourages users to enter stronger passwords? Moreover, are there text entry methods that are significantly worse in this respect and should not be permitted at all?

\section*{Acknowledgments}

This material is based upon work supported by the National Science Foundation under Grant Number 1228777. Any opinions, findings, and conclusions or recommendations expressed in this material are those of the author(s) and do not necessarily reflect the views of the National Science Foundation.

\balance
\bibliographystyle{IEEEtran}
\bibliography{passwordbib}

\begin{thebibliography}{10}
\providecommand{\url}[1]{#1}
\csname url@samestyle\endcsname
\providecommand{\newblock}{\relax}
\providecommand{\bibinfo}[2]{#2}
\providecommand{\BIBentrySTDinterwordspacing}{\spaceskip=0pt\relax}
\providecommand{\BIBentryALTinterwordstretchfactor}{4}
\providecommand{\BIBentryALTinterwordspacing}{\spaceskip=\fontdimen2\font plus
\BIBentryALTinterwordstretchfactor\fontdimen3\font minus
  \fontdimen4\font\relax}
\providecommand{\BIBforeignlanguage}[2]{{%
\expandafter\ifx\csname l@#1\endcsname\relax
\typeout{** WARNING: IEEEtran.bst: No hyphenation pattern has been}%
\typeout{** loaded for the language `#1'. Using the pattern for}%
\typeout{** the default language instead.}%
\else
\language=\csname l@#1\endcsname
\fi
#2}}
\providecommand{\BIBdecl}{\relax}
\BIBdecl

\bibitem{Herley:2012:RAA:2360743.2360824}
C.~Herley and P.~van Oorschot, ``A research agenda acknowledging the
  persistence of passwords,'' \emph{IEEE Security and Privacy}, vol.~10, no.~1,
  pp. 28--36, jan 2012.

\bibitem{cellinternet2013}
M.~Duggan and A.~Smith, ``Cell internet use 2013,'' sep 2013, pew Research
  Centers Internet \& American Life Project.

\bibitem{MacKenzie:2007:TES:1296062}
I.~S. MacKenzie and K.~Tanaka-Ishii, \emph{Text Entry Systems: Mobility,
  Accessibility, Universality}.\hskip 1em plus 0.5em minus 0.4em\relax San
  Francisco, CA, USA: Morgan Kaufmann Publishers Inc., 2007.

\bibitem{zhai2002performance}
S.~Zhai, M.~Hunter, and B.~A. Smith, ``Performance optimization of virtual
  keyboards,'' \emph{Human--Computer Interaction}, vol.~17, no. 2-3, pp.
  229--269, 2002.

\bibitem{norman1982alphabetic}
D.~A. Norman and D.~Fisher, ``Why alphabetic keyboards are not easy to use:
  Keyboard layout doesn't much matter,'' \emph{Human Factors: The Journal of
  the Human Factors and Ergonomics Society}, vol.~24, no.~5, pp. 509--519,
  1982.

\bibitem{gentner1983acquisition}
D.~R. Gentner, ``The acquisition of typewriting skill,'' \emph{Acta
  Psychologica}, vol.~54, no.~1, pp. 233--248, 1983.

\bibitem{goel2012walktype}
M.~Goel, L.~Findlater, and J.~Wobbrock, ``Walktype: using accelerometer data to
  accomodate situational impairments in mobile touch screen text entry,'' in
  \emph{Proc. of CHI'12}, 2012.

\bibitem{oulasvirta2013improving}
A.~Oulasvirta, A.~Reichel, W.~Li, Y.~Zhang, M.~Bachynskyi, K.~Vertanen, and
  P.~O. Kristensson, ``Improving two-thumb text entry on touchscreen devices,''
  in \emph{Proc. of CHI'13}, 2013.

\bibitem{bonneauPINs}
J.~Bonneau, S.~Preibusch, and R.~Anderson, ``A birthday present every eleven
  wallets? the security of customer-chosen banking {PINs},'' in \emph{Proc. of
  FC'12}, 2012.

\bibitem{azenkot2012touch}
S.~Azenkot and S.~Zhai, ``Touch behavior with different postures on soft
  smartphone keyboards,'' in \emph{Proc. of CHI'12}, 2012.

\bibitem{salthouse1986perceptual}
T.~A. Salthouse, ``Perceptual, cognitive, and motoric aspects of transcription
  typing.'' \emph{Psychological bulletin}, vol.~99, no.~3, p. 303, 1986.

\bibitem{Shannon:2001:MTC:584091.584093}
C.~E. Shannon, ``A mathematical theory of communication,'' \emph{SIGMOBILE Mob.
  Comput. Commun. Rev.}, 2001.

\bibitem{394764}
J.~Massey, ``Guessing and entropy,'' in \emph{Information Theory, 1994.
  Proceedings., 1994 IEEE International Symposium on}, 1994.

\bibitem{Burr:2011:SEA:2206278}
W.~E. Burr, D.~F. Dodson, E.~M. Newton, R.~A. Perlner, W.~T. Polk, S.~Gupta,
  and E.~A. Nabbus, ``Sp 800-63-1. electronic authentication guideline,''
  National Institute of Standards \& Technology, Gaithersburg, MD, United
  States, Tech. Rep., 2011.

\bibitem{Castelluccia:2012:NDSS:markov}
C.~Castelluccia, M.~D\"{u}urmuth, and D.~Perito, ``Adaptive password-strength
  meters from markov models,'' in \emph{Proc. of NDSS'12}, 2012.

\bibitem{Biddle:2012:GPL:2333112.2333114}
R.~Biddle, S.~Chiasson, and P.~Van~Oorschot, ``Graphical passwords: Learning
  from the first twelve years,'' \emph{ACM Comput. Surv.}, 2012.

\bibitem{Wilkes:1975:TSC:540274}
M.~V. Wilkes, \emph{Time Sharing Computer Systems}.\hskip 1em plus 0.5em minus
  0.4em\relax New York, NY, USA: Elsevier Science Inc., 1975.

\bibitem{Saltzer:1974:PCI:361011.361067}
J.~H. Saltzer, ``Protection and the control of information sharing in
  multics,'' \emph{Commun. ACM}, 1974.

\bibitem{Morris:1979:PSC:359168.359172}
R.~Morris and K.~Thompson, ``Password security: a case history,'' \emph{Commun.
  ACM}, 1979.

\bibitem{benefitspasswords}
M.~Jakobsson and M.~Dhiman, ``The benefits of understanding passwords,'' in
  \emph{Proc. of HotSec'12}, 2012.

\bibitem{Florencio:2007:LSW:1242572.1242661}
D.~Florencio and C.~Herley, ``A large-scale study of web password habits,'' in
  \emph{Proc. of WWW'07}, 2007.

\bibitem{Grawemeyer:2011:UMM:1994007.1994160}
B.~Grawemeyer and H.~Johnson, ``Using and managing multiple passwords: A week
  to a view,'' \emph{Interact. Comput.}, vol.~23, no.~3, 2011.

\bibitem{Yan:2004:PMS:1024867.1025014}
J.~Yan, A.~Blackwell, R.~Anderson, and A.~Grant, ``Password memorability and
  security: Empirical results,'' \emph{IEEE Security and Privacy}, 2004.

\bibitem{Kuo:2006:HSM:1143120.1143129}
C.~Kuo, S.~Romanosky, and L.~F. Cranor, ``Human selection of mnemonic
  phrase-based passwords,'' in \emph{Proc. of SOUPS'06}, 2006.

\bibitem{Ur:2012:YPM:2362793.2362798}
B.~Ur, P.~G. Kelley, S.~Komanduri, J.~Lee, M.~Maass, M.~L. Mazurek, T.~Passaro,
  R.~Shay, T.~Vidas, L.~Bauer, N.~Christin, and L.~F. Cranor, ``How does your
  password measure up? the effect of strength meters on password creation,'' in
  \emph{Proc. of USENIX Security'12}, 2012.

\bibitem{Egelman:2013:MPG:2470654.2481329}
S.~Egelman, A.~Sotirakopoulos, I.~Muslukhov, K.~Beznosov, and C.~Herley, ``Does
  my password go up to eleven?: the impact of password meters on password
  selection,'' in \emph{Proc. of CHI'13}, 2013.

\bibitem{Shay:2010:ESP:1837110.1837113}
R.~Shay, S.~Komanduri, P.~G. Kelley, P.~G. Leon, M.~L. Mazurek, L.~Bauer,
  N.~Christin, and L.~F. Cranor, ``Encountering stronger password requirements:
  user attitudes and behaviors,'' in \emph{Proc. of SOUPS'10}, 2010.

\bibitem{Weir:2010:TMP:1866307.1866327}
M.~Weir, S.~Aggarwal, M.~Collins, and H.~Stern, ``Testing metrics for password
  creation policies by attacking large sets of revealed passwords,'' in
  \emph{Proc. of CCS'10}, 2010.

\bibitem{Mazurek:2013:MPG:2508859.2516726}
M.~L. Mazurek, S.~Komanduri, T.~Vidas, L.~Bauer, N.~Christin, L.~F. Cranor,
  P.~G. Kelley, R.~Shay, and B.~Ur, ``Measuring password guessability for an
  entire university,'' in \emph{Proc. of CCS '13}, 2013.

\bibitem{Houshmand:2012:BBP:2420950.2420966}
S.~Houshmand and S.~Aggarwal, ``Building better passwords using probabilistic
  techniques,'' in \emph{Proc. of ACSAC'12}, 2012.

\bibitem{Inglesant:2010:TCU:1753326.1753384}
P.~G. Inglesant and M.~A. Sasse, ``The true cost of unusable password policies:
  password use in the wild,'' in \emph{Proc. of CHI'10}, 2010.

\bibitem{BroSas2003}
S.~Brostoff and A.~Sasse, ``{``Ten strikes and you're out''}: Increasing the
  number of login attempts can improve password usability,'' in \emph{Proc. of
  CHI'03 Workshop on HCI and Security Systems}, 2003.

\bibitem{Zviran:1999:PSE:1189462.1189470}
M.~Zviran and W.~J. Haga, ``Password security: an empirical study,'' \emph{J.
  Manage. Inf. Syst.}, 1999.

\bibitem{Chiasson:2009:MPI:1653662.1653722}
S.~Chiasson, A.~Forget, E.~Stobert, P.~C. van Oorschot, and R.~Biddle,
  ``Multiple password interference in text passwords and click-based graphical
  passwords,'' in \emph{Proc. of CCS'09}, 2009.

\bibitem{Wiedenbeck:2005:PDL:1090412.1090418}
S.~Wiedenbeck, J.~Waters, J.-C. Birget, A.~Brodskiy, and N.~Memon,
  ``Passpoints: design and longitudinal evaluation of a graphical password
  system,'' \emph{Int. J. Hum.-Comput. Stud.}, 2005.

\bibitem{Fahl:2013:EVP:2501604.2501617}
S.~Fahl, M.~Harbach, Y.~Acar, and M.~Smith, ``On the ecological validity of a
  password study,'' in \emph{Proc. of SOUPS'13}, 2013.

\bibitem{Schaub:2012:PEU:2406367.2406384}
F.~Schaub, R.~Deyhle, and M.~Weber, ``Password entry usability and shoulder
  surfing susceptibility on different smartphone platforms,'' in \emph{Proc. of
  MUM '12}, 2012.

\bibitem{Haque:2013:PIT:2516760.2516767}
S.~M.~T. Haque, M.~Wright, and S.~Scielzo, ``Passwords and interfaces: Towards
  creating stronger passwords by using mobile phone handsets,'' in \emph{Proc.
  of SPSM '13}, 2013.

\bibitem{rethinkingpasswords}
M.~Jakobsson and R.~Akavipat, ``Rethinking passwords to adapt to constrained
  keyboards,'' in \emph{Proc. of MoST}, 2012.

\bibitem{mohammadboth}
M.~Mannan and P.~V. Oorschot, ``Passwords for both mobile and desktop
  computers: Obpwd for firefox and android,'' \emph{;login}, no.~4, 2012.

\bibitem{hart:tlx}
S.~G. Hart and L.~E. Staveland, ``Development of nasa-tlx (task load index):
  Results of empirical and theoretical research,'' \emph{Advances in
  psychology}, 1988.

\bibitem{mentalrotation}
A.~Johnson, ``The speed of mental rotation as a function of problem-solving
  strategies,'' \emph{Perceptual and Motor Skills, 71, 803-806.}, Dec 1990.

\bibitem{Bonneau:2012:SMI:2437647.2437657}
J.~Bonneau, ``Statistical metrics for individual password strength,'' in
  \emph{International Workshop on Security Protocols}, 2012.

\bibitem{Kelley:2012:GAM:2310656.2310715}
P.~G. Kelley, S.~Komanduri, M.~L. Mazurek, R.~Shay, T.~Vidas, L.~Bauer,
  N.~Christin, L.~F. Cranor, and J.~Lopez, ``Guess again (and again and again):
  Measuring password strength by simulating password-cracking algorithms,'' in
  \emph{Proc. of SS\&P'12}, 2012.

\bibitem{Bonneau:2012:SGA:2310656.2310721}
J.~Bonneau, ``The science of guessing: Analyzing an anonymized corpus of 70
  million passwords,'' in \emph{Proc. of SP'12}, 2012.

\bibitem{johntheripper}
\BIBentryALTinterwordspacing
``John the ripper.'' [Online]. Available: \url{http://www.openwall.com/john/}
\BIBentrySTDinterwordspacing

\bibitem{hashcatsoftware}
\BIBentryALTinterwordspacing
``hashcat.'' [Online]. Available: \url{http://hashcat.net/hashcat/}
\BIBentrySTDinterwordspacing

\bibitem{outpost9}
\BIBentryALTinterwordspacing
outpost9, ``A list of popular password caracking wordlist,'' 2005. [Online].
  Available: \url{http://www.outpost9.com/files}
\BIBentrySTDinterwordspacing

\bibitem{facebookindex2010}
\BIBentryALTinterwordspacing
``Return of the facebook snatchers,'' 2010. [Online]. Available:
  \url{https://blog.skullsecurity.org/2010/return-of-the-facebook-snatchers}
\BIBentrySTDinterwordspacing

\bibitem{jtrmodifiedsingle2010}
\BIBentryALTinterwordspacing
``Optimizing john the ripper's ``single" mode for dictionary attacks,'' 2010.
  [Online]. Available:
  \url{http://reusablesec.blogspot.com/2010/04/optimizing-john-rippers-single-mode-for.html}
\BIBentrySTDinterwordspacing

\bibitem{Weir:2009:PCU:1607723.1608146}
M.~Weir, S.~Aggarwal, B.~d. Medeiros, and B.~Glodek, ``Password cracking using
  probabilistic context-free grammars,'' in \emph{Proc. of SP '09}, 2009.

\bibitem{doi:10.1080/14786440009463897}
K.~Pearson, ``On the criterion that a given system of deviations from the
  probable in the case of a correlated system of variables is such that it can
  be reasonably supposed to have arisen from random sampling,''
  \emph{Philosophical Magazine Series 5}, 1900.

\bibitem{doi:10.2307/2340521}
R.~Fisher, ``On the interpretation of 2 from contingency tables, and the
  calculation of p,'' \emph{Journal of the Royal Statistical Society}, 1922.

\bibitem{isbn:9781446200469}
A.~Field, J.~Miles, and Z.~Field, \emph{Discovering Statistics Using R}.\hskip
  1em plus 0.5em minus 0.4em\relax SAGE Publications Ltd, 2012.

\bibitem{barbarataba2007}
B.~Tabachnick and L.~Fidell, \emph{Using Multivariate Statistics}.\hskip 1em
  plus 0.5em minus 0.4em\relax Boston, MA, USA: Pearson, 2007.

\bibitem{Stevens:1986:AMS:21800}
J.~Stevens, \emph{Applied Multivariate Statistics for the Social
  Sciences}.\hskip 1em plus 0.5em minus 0.4em\relax Hillsdale, NJ, USA: L.
  Erlbaum Associates Inc., 1986.

\end{thebibliography}

\end{document}